\documentstyle [11pt,psfig]{article}
\begin{document}
\title{Luminosity Distributions within Rich Clusters - II: \\
Demonstration and Verification via Simulation}
\author{Simon P. Driver, Warrick J. Couch\\
School of Physics, University of New South Wales,\\ 
Sydney, NSW 2052, Australia\\
Steven Phillipps, \\
Astrophysics Group, Dept. of Physics, University of Bristol, \\
Tyndall Avenue, Bristol BS8 1TL, UK, \\
and, \\
Rodney Smith, \\
Dept Physics and Astronomy, University of Wales College of Cardiff,\\
PO Box 913, Cardiff CF2 3TH, UK \\}

\maketitle

\begin{abstract}
We present detailed simulations of long exposure CCD images.
The simulations are used to explore the validity of the statistical method for
reconstructing the luminosity distribution of galaxies within a rich cluster
{\it i.e.} by the subtraction of field number-counts from those of a 
sight-line through the cluster. In particular we use the simulations to 
establish the reliability of our observational data presented in Paper 3.
Based on our intended CCD field-of-view (6.5 by 6.5 arcmins) and a 1-$\sigma$
detection limit of 26 mags per sq arcsecond, we conclude that the luminosity 
distribution can be robustly determined over a wide range of absolute 
magnitude ($-23 < M_{R} < -16$) provided:

(a) the cluster has an Abell richness 1.5 or greater, 

(b) the cluster's redshift lies in the range $0.1 < z < 0.3$, 

(c) the seeing is better than FWHM 1.25$''$ and 

(d) the photometric zero points are accurate to within
$\Delta m = \pm 0.12$. 

If these conditions are not met then the recovered luminosity distribution
is unreliable and potentially grossly miss-leading. Finally although the 
method clearly has limitations, within these limitations the technique 
represents an extremely promising probe of galaxy evolution and environmental 
dependencies.

\noindent
{\bf Keywords:} galaxies: luminosity function, mass function - 
galaxies:evolution.
\end{abstract}

\section{Introduction}
In the first paper in this series (Smith, Driver \& Phillipps 1997; hereafter 
Paper I), we compared the derived luminosity distributions (LD)\footnote{We 
adopt the nomenclature of using luminosity distribution (LD) to refer to an 
observed distribution and luminosity function (LF) when describing either an 
analytic fitted function to this distribution or the input function to our
simulations. Here we use the analytical form of the Schechter function 
throughout (Schechter 1976).} of three
rich clusters, A963 (Driver {\it et al.} 1994), A2554 (Paper I) and Coma
(Bernstein {\it et al.} 1996), and postulated the existence of a
ubiquitous dwarf-rich LD for rich clusters. This has had
additional support from results published by other groups, in particular 
Wilson {\it et al.} (1997) who applied similar methods to recover the LDs
in A1689 \& A665, and found similar dwarf-rich LFs.
These results are based on a statistical method which 
relies on the subtraction of galaxy number-counts detected in a non-cluster 
sight-line from those detected in a cluster sight-line. The observed 
number-count excess along the cluster's sight-line is a direct representation 
of the cluster population. The method is observationally simple, requires 
minimal telescope time, yet provides a direct measurement of the fundamental
luminosity density-distribution of galaxies within that cluster. This is only 
otherwise achievable via a time-intensive spectroscopic survey. Furthermore, 
even if telescope time limitations are neglected, spectroscopic surveys can 
only probe to low luminosity levels in nearby groups (e.g. de Propris {\it et 
al.} 1995) and nearby clusters (see for example Bernstein {\it et al.} 1996; 
and Bivanio {\it et al.} 1996). To assemble a sufficiently large number of 
measurements of the luminosity distribution of galaxies over a range of 
environments and epochs a statistical approach is required. This is the goal
of these next two papers in the series, published here back-to-back. In this
paper (Paper II), we first fully assess the accuracy, limitations and the
presence of any systematic dependencies in the method via extensive and
realistic simulations. In the following paper (Driver, Couch \& Phillipps
1998; hereafter 
Paper III), we present new measurements of the LD for a sample of 7 rich 
clusters, armed with a very firm understanding of the characteristics 
and quality of these data from our simulations.  

The plan of this paper is as follows: 
In \S 2 we describe in detail the process of simulating individual galaxies, 
the range of types, the magnitude-density calculation and the simulation of 
noise processes. In \S 3 we discuss our chosen photometry package 
(SExtractor, Bertin \& Arnout 1996). In \S 4 we carry out our simulation of 
galaxy clusters and in \S 5 we run extensive simulations to explore the 
reliability of the technique and the range of observable parameters over which 
our LD recovery technique is valid. The final section (\S 6) contains an 
overall summary of our findings.

\section{The Simulation Process}

The principal aim of the simulation software is to adopt the main aspects of 
galaxy parameterisation to generate realistic deep CCD images. Below we define
our adopted generic types, our method for the determination of the 
line-of-sight number counts for each type, and the process of simulation. 

\subsection{Simulated morphological types}

To create a realistic CCD image we adopt four basic generic galaxy types
selected to span the range from well
ordered early systems to asymmetrical late systems. These types are: 
Ellipticals (E), spirals (Sb), irregulars (Irr) and dwarf ellipticals (dE). 
The field LFs and luminosity-surface brightness relations for 
these types are taken from Marzke {\it et al.} 
(1995) and Binggeli (1994) respectively (c.f. Table 1). In addition 
to the four generic galaxy types, stars are also simulated. The details of each
generic type are as follows (see also Table 1) :

\vspace{0.5cm}

\noindent
\begin{description}

\item {\bf Ellipticals:}
These are assumed to exhibit $r^{\frac{1}{4}}$ 
de Vaucouleurs (1959) light profiles. Their 
ellipticity (or $(a-b)/a$, where $a$ and $b$ are the major and minor axis), 
is randomly distributed within the range 0.0 to 0.7 ({\it i.e.} E0---E7 in 
conventional nomenclature) with equal probability. The light-profile is 
simulated out to $\sim 6$ half-light radii. 

\item
{\bf Spirals:}
The majority of the total flux (75\%) is simulated as a perfect 
exponential disk (Freeman 1970) with inclination in the range 0---90 degrees. 
The remaining 25\% of flux is placed in a centrally located bulge (profiled as 
an elliptical galaxy as discussed above). This bulge-to-disk ratio is 
consistent with a stage 3 galaxy (Sb) on the 14 step revised Hubble 
classification scheme (Simien \& de Vaucouleurs 1986). 
The inclination angle ($i$) and position angle (PA) of the disk are
randomly generated with equal probability. The ellipticity is derived from
the inclination ($i$) and intrinsic axis ratio ($Q_{o}$) as follows
(c.f. Hubble 1926):
\begin{equation}
\cos^{2}(i)=\frac{(\frac{a}{b})^{2}-Q_{o}^{2}}{1-Q_{o}^{2}}
\end{equation}
where,
$Q_{o}$, the intrinsic axis ratio, is taken as 0.2 (c.f. Holmberg 1946).
The disk is profiled out to 10 scale-lengths. 
Galaxy disks are treated as optically thin, and therefore the
simulated galaxy's total magnitude 
remains independent of inclination (but surface brightness does not).

\item
{\bf Irregulars:} The irregulars are modeled on a disk system containing a
number of active star-forming regions. To achieve this 20\% to 50\% of the 
total luminosity is distributed as an exponential disk, identical to the 
mid-type spirals. The remaining flux is randomly distributed within three scale
lengths across the primary disk as secondary superimposed mini-disks. An 
intrinsic disk axis ratio of $Q_{o}$ = 0.4 is adopted, implying a minimally 
rotating thick disk. Each simulated HII complex contains 5\% of the total 
integrated flux. The central surface brightness of these HII complexes is 
randomized over a range of 2 mags arcsec$^{-2}$ about the central surface 
brightness of the underlying disk.

\item
{\bf Dwarf Ellipticals:}
These systems are numerous within cluster environments (c.f. Ferguson \&
Binggeli 1994) and are simulated here as perfect exponential disks with no
bulge or irregularity. An intrinsic disk axis ratio of
$Q_{o}$ = 0.7 is adopted to reflect the lack of strong dynamical rotation
and the probable tri-axial nature of these systems.

\item
{\bf Stars:} These are added after convolution with the Gaussian seeing disk
and are simulated as a pure Gaussian profile, {\it i.e.} the diffraction 
rings of the Airy disk are neglected. For those stars with excessive flux, 
this excess is 
redistributed into two diffraction spikes aligned along the axis of the frame 
edges and a general low surface brightness ``halo'' is superimposed. The
intention is to create a more realistic image rather than accurate
treatment of bright stars, as few will be simulated per field of 
view and, at these magnitudes, the star/galaxy separation is trivial.
Typically in the simulations described here the ``diffraction spiking'' occurs 
at magnitudes of $m_{R} < 18.5$.
\end{description}

\subsection{Simulating a field sight-line}
To generate a simulated field sight-line we adopt the numerical 
morphological number-count model of Driver {\it et al.} (1995). 
This model is based on the following: an observed field 
luminosity function for each galaxy type (c.f. Marzke {\it et al.} 1995),
the visibility---distance relationships for a standard flat cosmology (c.f. 
Phillipps, Davies \& Disney 1990), K-corrections based on data from
Pence (1976); King \& Ellis (1985) and the observed luminosity---surface 
brightness relationships for each of the four adopted galaxy types (taken from 
the schematic of Binggeli 1994). The exact values used are summarized in 
Table 1 and more extensive details with regards the generation of differential
galaxy number-counts are described in Driver {\it et al.} (1994); Driver, 
Windhorst \& Griffiths (1995) and Driver {\it et al.} (1995). The only 
adjustable parameters in this prescription are the normalisation of the 
individual luminosity functions, and these have been
optimised such that our simulated morphological number-counts match those 
observed in the Hubble Deep Field (c.f. Williams {\it et al.} 1996 and other 
deep Hubble Space Telescope WFPC2 images (c.f. Odewahn {\it et al.} 1996).

To allow for field-to-field variation the final numbers of each galaxy type at 
each magnitude interval are adjusted to reflect Poisson counting statistics.
In those cases where less than one object is predicted ({\it e.g.} at bright 
magnitudes) the number-density is taken as a probability of that object 
occurring. The density of galaxies is generated in 0.25 magnitude intervals 
from $m_{R} = 14.0$ to $m_{R} = 26.0$. 

No attempt was made to simulate field-to-field clustering or lensing (either
weak field lensing or strong cluster lensing). This decision was taken to 
limit the complexity of the software. We note thogh that Paper III does 
explore the observed field-to-field variation for our 7 non-cluster sight-lines
and find that variation is close to Poisson expectation over all magnitudes.
With regards to lensing we note firstly, 
that the slope of the  background number-counts is close to 0.4 (where the 
effects of amplification and volume narrowing cancel out, Broadhurst et al 
1998), and secondly that the detailed treatment by Trentham (1998) for A665 
and A963 implied that the lensing correction is small except at very small 
radii ($< 200$kpc) where a 10\% increases of the background population might 
be expected. Hence for very small cluster cross-sections lensing could become
an issue.

With the catalogue defined, we construct a data frame by 
simulating light profiles, orientations and positions followed by the various
noise processes. This sequence is summarised below:

\begin{description}

\item{\bf Galaxy Simulation:} Ellipticals, mid-type spirals, irregulars and
dwarf ellipticals are simulated in order of apparent magnitude from 
brightest to faintest magnitudes. Orientations, positions and
inclinations are generated randomly on the fly. A running tally is kept of the 
assigned flux for each object and any residual positive or negative flux, 
due to the integer pixel size, is re-allocated in two possible ways. For small
discrepancies the central pixel value is adjusted; for larger 
discrepancies\footnote{Occasionally two galaxies will overlap such that a 
single pixel may ``flood'', in which case the flux of the later galaxy is 
reallocated to the nearest available pixel. This will cause occasional 
oddities mostly erased by the later seeing disk
convolution. However it ensures that all 
the allocated flux is assigned as near as possible to the object and all 
photometric errors are due to noise fluctuations and photometric 
error/assumptions rather than the simulations. {\it i.e.} flux allocations are
ALWAYS correct with the occasional compromise in profile to accommodate this.}
the residual flux is overlaid as a Gaussian core around the central pixel value
({\it i.e.} a variable-sized compact nucleus).

\item{\bf Object Noise Added:} At this point the simulated image contains 
perfect profiles on a uniform background. To simulate shot
noise, each pixel value in the image is randomly re-allocated according to
Poisson statistics. Hence, from this point on an 
individual object's flux is no-longer absolutely exact but reflects the real 
Poissonian variation expected in an object's signal.

\item{\bf Gaussian Convolution:} The IRAF routine GAUSS is used to convolve the
image with a perfectly symmetrical Gaussian seeing disk of the desired
full width half-maximum (FWHM). For compact objects the seeing convolution acts
to smooth the image somewhat diminishing the effect of the shot noise process.
The point spread function is assumed perfectly symmetrical over the entire
image f.o.v., with telescope tracking perfect and optics immaculate.

\item{\bf Stars Added:} Stars are profiled with Gaussian distributions using 
the same FWHM as in the previous step. Stars with excessive flux have this 
excess redistributed as simulated diffraction spikes and an underlying
low surface brightness disk. Magnitudes are once again forced to be correctly 
allocated regardless of the stars brightness and the amount of 
``flooding/blooming''. Shot noise is not simulated for the stars as the 
photometric accuracy and detectability of stars is not under investigation.

\item{\bf Sky Noise Added:} A random number generator is used to produce
simulated sky noise with a perfect Gaussian distribution. Given that the
galaxies are simulated to fainter levels than the detection limit,
the final noise distribution is marginally skewed to positive fluxes, as
occurs in real data.
\end{description}

~

\noindent
Figure 1 shows a montage of six images, illustrating the development of a
simulated image through the stages previously outlined. The panels from
top to bottom and left to right show a 1 by 1 arcmin section of a simulated
cluster image at various stages: (a)\,simulated field galaxies, all types;
(b)\,cluster galaxies added; (c)\,with shot noise added (barely noticeable at 
the grey levels chosen); (d)\,after convolution with a 0.9$''$ FWHM Gaussian 
``seeing'' disk; (e)\,with stars added; and finally, (f)\,with sky noise 
added. Note in particular the fattening and shortening of edge--on disks,
the loss of faint objects and the loss of low surface brightness objects as
these various processes are applied. 

\section{Detection Software}
Our chosen object detection and photometry package is
SExtractor described in detail by Bertin \& Arnout (1996). The finding 
algorithm used by SExtractor searches for galaxies with a minimum number of 
connected pixels above a specified background after convolution with a preset
filter; in this case a $3\times3$ top-hat filter.
Each detection is then re-analysed and a 
number of detailed measurements made, including: corrected isophotal 
magnitude, Kron magnitude, isophotal area and a star/galaxy estimation. Note 
that the SExtractor package adopts Kron magnitudes when the object is isolated
and uses the corrected isophotal magnitude for crowded regions. 
In addition SExtractor has a logical strategy for 
dividing blended objects into components based on the ratio between the 
sub-peaks and the total flux. 

Kron magnitudes are used for estimating total fluxes and are based on the Kron 
radius which is defined as: 
$R_{K}=\frac{\sum I(r) r \delta r}{\sum I(r) \delta r}$ (Kron 1978). The
magnitude is normally measured through an aperture of radius $r = 2.5 R_{K}$, 
which is that recommended for the inclusion of $> 90$\% of 
the flux for ellipticals and spirals. To test how the irregulars faired we 
measured the flux for random samples of 54 galaxies of each type with 
constant magnitude of $m_{R}=18.0$ and varied the aperture radius from 
2.0--5.0\,$R_{K}$. Table 2 shows the resulting flux returned based on the 
following specific types: Ellipticals with $M_{R} = -22.0$ mags, 
($\mu_{o} = -13.0$ mags per sq arcsec); 
mid-type spirals with $M_{R} = -20.0$ mags,
($\mu_{o} = -19.3$ mags per sq arcsec); 
irregulars with $M_{R} = -17$ mags, 
($\mu_{o} = -21.7$ mags per sq arcsec); 
and dwarf ellipticals with $M_{R} = -18.0$ mags,
($\mu_{o} = -19.0$ mags per sq arcsec).
On the basis of these tests a more 
conservative radius of $r = 3.5 R_{K}$ was 
selected such that our photometric error is less than the typical 
magnitude zero point error 
expected in our data ($\Delta m = \pm 0.05$).
As a further test of both the galaxy simulations and the SExtractor software
we simulated $\sim 4000$ galaxies of each type ranging from $m_{R} = 18$ to
$m_{R} = 26$ and added noise equivalent to $m_{R} = 26.0$ mags arcsec$^{-2}$.

Figure 2 shows the resulting median magnitude difference (errorbars
are the upper and lower quartiles) versus the known simulated magnitude, for 
the five object types. Also shown in Figure 2 are the theoretical
upper and lower quartiles for a perfect point source (star) convolved with the
simulated Gaussian sky noise fluctuations.
From Figure 2 we can see that the fluxes are both accurately allocated and
measured, regardless of galaxy type and orientation over bright magnitudes. 
The discrepancy for the brightest stars is due to the simulated diffraction 
spikes which extend outside of the $r = 3.5 R_{K}$ aperture and are present 
for $m_{R} < 18.5$ only. For all galaxy types magnitudes become unreliable 
($\Delta m >$ 0.1) at $m_{R} \approx 24$ and totally unreliable fainter than 
$m_{R} \approx 24.5$; note that this is 1.5 mags above the theoretical 
$1\sigma$ point source detection limit! 
Figure 3 shows the corresponding fractional completeness as a function of 
input magnitude. It can be seen that incompleteness becomes significant
at $m_{R} \approx 24.0$; the completeness drops to 90\% completeness at 
$m_{R}=23.95$, $m_{R}=24.3$, $m_{R}=24.0$ and $m_{R}=24.4$ for ellipticals,
mid-type spirals, irregulars and dwarf ellipticals, respectively. Stars fall 
below the 90\% completeness limit beyond $m_{R}=24.25$.
That the ellipticals are 'lost' first is a testimony to the power of
cosmological
surface brightness-dimming or the ``cosmic guillotine'' effect (see Koo \& Kron
1993).

\section{Cluster Simulation}

\subsection{Cluster type}
There are typically three major classifications of cluster type as listed in 
Oemler (1974): spiral-rich, cD-dominated or spiral-poor. These three 
different types of cluster are associated with an irregular or flat profile, 
fully viralised radial profile (1/$r$) or a semi-relaxed intermediary profile, 
respectively. 
A strong relation between morphology and local projected galaxy density is also
seen across all 3 cluster types with a smooth transition between an 
E/S0-dominated mix in the higher density regions to one which is 
spiral-dominated in the lower density regions (Dressler 1980).
In the regular cD and spiral-poor clusters, this results in an elliptical
distribution which is much more centrally concentrated than that for spirals.
To accommodate this, we adopt two principle profile shapes. These two profiles 
are: a purely radial 1/$r$ distribution giving a centrally concentrated 
population (consistent with the distribution of ellipticals), and, 
a Gaussian distribution where the FWHM is 1 Mpc or 1/3 of the Abell radius
(consistent with the less concentrated spiral distribution).
Hence to simulate a cD or spiral poor cluster we require a cluster with a
high density of
ellipticals distributed in a (1/$r$) profile and a low density of spirals
distributed in a Gaussian. The net result for the overall profile is a slightly
extended (1/$r$) profile consistent with the formulation of a cD cluster as
presented in Oemler (1974). Similarly for a spiral-rich cluster we have an 
equal density of ellipticals and spirals resulting in an overall profile 
with a predominantly Gaussian distribution again consistent with
the specification of Oemler (1974) {\it i.e.} by changing the morphological
mix we are also changing the cluster profile shape.
Figure 4 shows the two profiles (left side; elliptical (1/$r$) profile (upper) 
and spiral Gaussian profile (middle)) and the equivalent projected density 
distributions (right side). 
A practical modification is made to the (1/$r$) profile
by randomising positions within the central 100\,kpc as
a purely (1/$r$) profile produces an unrealistic number of galaxies at the
cluster centre. 

\subsection{The Reference Cluster}
For the simulations presented in this and the next section, it was convenient 
to define a reference cluster. The strategy is that once the reference 
cluster has been defined we can then explore deviations from this parameter 
space rather than simulate all possible clusters (which for obvious reasons is 
impractical). We model our reference cluster on the
parameter space for which we have obtained most of our data, {\it i.e.}
spiral-poor richness 3 clusters at redshift $\sim 0.15$.
Hence we adopt the cluster profiles for ellipticals and spirals as discussed
above with a morphological mix ratio of E/S0:Sabc of 1:1\footnote{This is the
morphological mix within the full Abell radius, within the central core only, 
{\it i.e.} $r < 500$\,kpc, the mix is $\sim 2:1$ consistent with that of a 
spiral-poor cluster, see Figure 4 lower left panel}. The lower panels in  
Figure 4 show the morphological mix as defined within some
fraction of the Abell radius (left) and the combined projected density
distribution of ellipticals and spirals (right). For the dwarf 
ellipticals, we found in Paper I that the lower luminosity population is
much less strongly clustered (even after consideration of the difficulty of 
detecting dwarfs in the cluster core); we therefore also select
a Gaussian profile distribution for this population.

Figure 5 shows an example simulation of our
reference cluster and the strong central clustering of ellipticals is clearly
apparent. A comparison field region has also been simulated (although not
shown) and the number counts for both images determined using the SExtractor
software as described earlier. 
The final counts are corrected for
the diminishing area available to fainter objects (see Paper I for details). 
Figure 6 shows the resulting number-counts and
here we compare the input and detected number-counts
for (a) the field, and (b) the cluster + a second independent field.
The lower panel of Figure 6 
shows the recovered LD for this cluster compared to the known input LD.
Note that in this single 
simulation, $\sim1200$ galaxies are detected in the field image and $\sim1500$ 
galaxies are detected in the cluster image ({\it i.e.} $\sim2700$ 
total galaxies detected out of $\sim5100$ simulated for the 
two sight-lines). 
The completeness limit ($m_{R} \approx 24$), as defined 
by the departure of the detected number-counts from the known input 
number-counts, is in agreement with the earlier photometric tests.
The recovered LD for our reference cluster is a good
representation of the input distribution demonstrating the qualitative
validity of the technique. In order to evaluate the match quantitatively we
apply a Kolmogorov-Smirnov Two-Sample Test. A two sample test is applied as
the input distribution is itself drawn from the initially
specified smooth LFs (as we are 
sampling only a portion of the cluster's population). 
The KS value is determined
in the normal manner, {\it i.e.} both distributions are normalised to unity
within the magnitude range of comparison and the cumulative distributions
compared, the largest discrepancy between the two distributions is then the 
KS value. To evaluate the 
probability of these two distributions being the same we convert the KS 
value to a $\chi^{2}$ value and determine the probability from the incomplete
gamma function. The conversion is  given by Wall (1996):

\begin{equation}
\chi^{2} = 4 (KS)^{2}\frac{N_{i} N_{o}} {N_{i}+N_{o}}, 
\end{equation}
where $N_{i}$ is the input number of cluster galaxies above our
magnitude cutoff and $N_{o}$ is the observed number of cluster galaxies above 
our magnitude cutoff. 

We now define a `goodness-of-fit' parameter, GOF, to be the 
evaluated $\chi^{2}$ probability of the two 
distributions being the same, expressed as a percentage. Hence for the single 
simulation shown in Figure 6 the GOF=94.7 \%. 
Since a single simulation may be unrepresentative due to a statistical 
vagary in this particular arrangement of stars and galaxies,
we repeat the measurement a number of times and 
estimate the mean GOF and its standard deviation.
Figure 7 shows the recovered LDs compared to the actual LDs and the individual 
GOFs for eleven simulations of our reference cluster. 
In all cases the agreement is good to excellent. (Note that a total of
22 sight-lines or $\sim 56,000$ galaxies were simulated). The
lower right panel of Figure 7 shows the average LDs; the mean overall GOF 
of the eleven simulations is 93 $\pm 4$ \%.

\section{The Simulations}
We now explore the validity of our technique for recovering a cluster's LD 
over a wider range of parameter space. To determine the optimal parameter 
values we adopt a reference cluster ($z=0.15$, Richness = 3, $\alpha=-1.0$, 
E:S0/Sabc mix = 1:1) 
and investigate deviations from this scenario by adjusting
one parameter at a time. In particular we wish to determine how far our 
technique for LD recovery can be applied in terms of cluster redshift
and cluster richness and to assess its dependence on observables such as 
seeing conditions and zero magnitude point errors. In each case we simulate
eleven realisations of each scenario over the range of
parameters we wish to explore.

These simulations require the specification of detector parameters
such as field-of-view (f.o.v.), pixel size, noise, exposure time etc. 
Here we model our simulations on the observational
configuration used in collecting our new data (Paper III): the f/3.3 
prime-focus of the 3.9\,m Anglo-Australian Telescope (AAT) equipped with
a Tektronix 1K$^{2}$ CCD, giving 0.39\,arcsec pixels, a field-of-view 
of $6.5\times 6.5$\,arcmin and an exposure time of 90 mins.

\subsection{The morphological mix}
We have already discussed the three principle cluster types which are
distinguished by the variation from dense, cD-dominated, spiral-deficient 
to the more loosely bound spiral-rich clusters. Hence it is clear that we
must consider the morphological mix and profile shape together. The definition
of our cluster profile is such that an entirely spiral-deficient 
cluster will have a (1/$r$) profile and a spiral-rich cluster a Gaussian  
profile. Hence 
by simply adjusting the morphological mix ratio of ellipticals to spirals
we explore, simultaneously, both the morphology and profile dependence of the 
technique. Figure 8a shows the GOF results, with each point calculated from 
the mean of 11 simulated 
clusters and field sight-lines. The simulations span the E/S0:Sabc mix range 
in steps from 4:1 to 1:1 (where the mix is defined within the Abell 
radius). The errorbars indicate the standard deviation of the
values (not of the mean {\it i.e.} we have not divided the errors by 
$\sqrt{N}$), so that they more reflect the range of GOFs from individual
values rather than the accuracy to which we have evaluated the mean GOF.
The flat GOF shows that the technique is insensitive (at these redshifts,
richnesses etc) to the morphological mix and profile shape. The former is
hardly surprisingly as at the faint limits morphological information
is lost while at brighter limits we do not expect any bias nor have we
detected any
in our photometric tests. The latter is perhaps more surprising as one might
expect the denser profile to be more confused; nevertheless the results
indicate this is not a significant concern.

\subsection{The faint end slope}
As stated previously,
much attention has recently been drawn to the faint end slope of the 
luminosity function, the clusters observed so far all exhibiting an
apparently universal dwarf-rich 
LD (Paper I). The question of whether the same form might also be true of the 
field LF is highly topical (Phillipps \& Driver 1995; Driver \& Phillipps 
1996). However, we first
need to ask whether the technique can reliably recover any shape of LD.
To explore this possibility we adopt the reference cluster and vary only the
faint slope parameter for the dwarf ellipticals, {\it i.e.} the $\alpha$ 
parameter in the adopted dE Schechter function (Schechter 1976). In this 
simulation we explored the range $-0.5 < \alpha < -2$ with $M_{R}^{*} = -19.0$
and $\phi_{dE}^{*}=\phi_{E}^{*}$. The results are shown
in Figure 8b and indicate that our technique yields an acceptably high GOF for
any faint end slope. The lowest GOF value of $85\%$ occurs for a steeply 
declining faint end LF slope. 
Results are marginally more reliable for the previously observed
values of $\alpha \sim -1.5$. These simulations therefore add significant 
weight to the results of Paper I.

\subsection{Cluster richness}
The Abell richness parameter varies within the ACO catalogue (Abell, Corwin \& 
Olowin 1989) from a value of 0 to 5. Richness is defined as 0, 1, 2, 3, 4 or
5 if the number of galaxies in the range $m_{3}$ to $m_{3}+2$ within the Abell 
radius is:
30---49, 50---79, 80---129, 130---199, 200---299, 300 or over, respectively,
where $m_{3}$ is the magnitude of the third brightest object in the cluster.
The main concern with the LD recovery technique is that it relies on the 
contrast of cluster members being significant in comparison to the typical 
field-to-field 
variation from one sight-line to the next. To examine this we simulated
clusters over the whole range of richness by adjusting the $\phi_{*}$ values of
the three cluster LFs. To reduce the coarseness of the richness class
scale we also devised half richness classes. Figure
8c shows the results and clearly cluster richness is of
paramount importance. The drop in GOF occurs rapidly for 
richnesses less than 1.5 suggesting there is little reliability in the 
reconstruction of the LDs of poor clusters. 
Going to a large format CCD, giving a larger f.o.v. is likely to partially
alleviate the problems.
In these simulations we have reduced the dwarf component in line 
with the luminous galaxies to preserve an overall flat LF. If the density of 
dwarfs is independent of giants we may have biased our results adversely.
The net result is that a richness limit of 1.5 is a realistic lower-limit
if the cluster LF is indeed flat in these environments, but which may need to
be re-evaluated if the dwarf and giant densities are unrelated.

\subsection{Cluster redshift}
If a cluster is too near then the physical extent seen by a contemporary
CCD will be small and consequently fewer cluster members will be seen. 
Conversely, at large distances, the cluster may become confusion limited, 
unresolved ({\it i.e.} 
cluster diameter smaller than f.o.v.) or simply too faint for our technique to 
work. Figure 8d shows the results from a set of simulations spanning the 
redshift range 0.0 to 0.5. The results indicate that both of these concerns are
valid. For the simulated f.o.v. chosen (6.5 by 6.5 arcmins$^{2}$) the 
optimal redshift for this technique appears to lie in the range 
$0.1 < z < 0.3$. At a completeness limit of $m_{R} = 24.0$ this 
equates to an absolute magnitude limit in the range $-15 < M_{R} < -17$ 
respectively. 
Wider f.o.v. detectors (mosaics or photographic emulsions) are likely to be 
more appropriate for
examining low redshift clusters although stellar and background confusion 
(Drinkwater {\it et al.} 1996) are more severe. 
At the high redshift end it is clear that compactness of the cluster core leads
to a severe confusion limit. However, it is likely 
that {\it Hubble Space Telescope} WFPC2 observations will extend the 
reliability of the technique to higher redshifts.

\subsection{Seeing}
We test the effect of the seeing in order to assess when the observing
conditions become too poor for our technique. Again we adopt the reference
cluster and Figure 8e shows the results of varying the seeing FWHM. As 
expected the GOF deteriorates badly in seeing worse than 2$''$ where 
the fainter galaxies are simply blended in with the bright ones or the mean 
surface brightness of compact objects falls below the detection threshold. 
The technique does seem to be almost independent of seeing over the
range 0$''$ to 1.5$''$. Surprisingly there is a slight tendency for $\sim1''$ 
seeing to give a better result than sub-arcsec seeing. This is likely 
to be due to two factors: The simulated pixel size and the over de-blending 
of irregulars by the SExtractor software. In these simulations, 0.39$''$ 
pixels have been used and in sub-arcsecond seeing, compact objects and noise
spikes become indistinguishable, raising the level of confusion. The second 
problem of de-blending is due to the complex structure of the simulated 
irregulars which in $\sim1''$ seeing are unresolved but in sub-arcsec 
seeing may be de-blended into several objects (c.f. Figure 1a). 
It would appear, therefore, that seeing of
FWHM $\sim 1''$ is optimal for providing the best GOF results.
Pixel size is also likely to be important, although we have not explored this
here. It is possible that the decline of the GOF in good seeing is related to
undersampling of the seeing disk. Of particular curiosity is the dramatic
dip in the GOF at a FWHM of 2.1$''$; the variation in the eleven simulations 
indicate that this is not a random error but is somehow systematic.

\subsection{Zero point error}
An alternative explanation to the turn-up seen in A963 (Driver {\it et al.}
1994a) is that there was a photometric zero point error between the cluster and
field sight-lines. There is some justification for this concern,
as the instrument with which the data were obtained,
``Hitchhiker'', was a parallel instrument for which calibration data were 
collected sporadically and instrument consistency was
relied upon. This criticism has been partly mitigated by recent observations of
other rich clusters with well calibrated data which 
have revealed similar LDs to A963 (c.f. Wilson {\it 
et al.} 1997). Nevertheless it seems prudent to ask how accurate the photometry
must be for the technique to be valid. Here we adjust the photometric zero 
magnitude point of firstly the cluster sight-line and then the field 
sight-line. The resulting GOF curve is 
not shown but the mean GOF was found to fall below 80\% if the photometric zero
points are in error by greater than $\Delta m = \pm 0.12$. (For a field number
count slope around 0.4 this implies a relative error of 12\% in numbers 
between sight lines).

\section{Conclusions}
The simulations presented here demonstrate the validity of applying a purely
statistical photometric method to reconstruct a cluster's Luminosity 
Distribution (LD). The 
simulations are used to generate detailed CCD images of field and cluster
sight-lines and incorporate factors such as: morphological types 
(E, Sb, Irr, dE \& Stars); morphological field luminosity functions;
surface-brightness luminosity dependencies; cosmological effects (surface 
brightness dimming, K-corrections, luminosity and angular distances); 
ellipticities, 
inclinations and disk axis-ratios; object overlap; cluster profiles (1/r, 
Gaussian), richness, redshift, luminosity function, morphological mix and
crowding; noise processes (shot-, sky-, unresolved objects); stellar 
contamination; atmospheric
seeing; detector and instrumental characteristics; image detection and 
photometry. Not simulated are: cosmic rays; CCD read-out noise, dark current 
and defects; and opacity (our simulations assume all objects are optically 
thin). Several thousand sight-lines were generated each containing several 
thousand objects with morphological number-counts matched to deep {\it Hubble 
Space Telescope} observations, images were created and galaxy catalogs 
generated in an identical manner to the data presented in Paper III. The 
derived cluster LDs were compared to the input distributions via a 
Kolmogorov-Smirnov statistical test to assess accuracy. As an aside we note 
that an accurate and complete faint galaxy catalogue can only be generated to 
a magnitude limit well above the point source detection limit. For a fully 
sampled seeing disk of $1''$ this limit is 2 magnitudes ({\it i.e.} for a 
point source detection limit of 26 mags the catalogue would only be complete 
and reliable to a limit of 24 mags). 

As the accuracy of the method is dependent on both the cluster properties
and the detector characteristics it is not
possible to define universally the cluster space over which the method is 
valid. Instead we demonstrate the specific cluster parameter space for our 
chosen detector through which the data presented in Paper III were collected.
Hence the simulations presented here were specifically to mimic a 90 minute 
exposure on the Anglo-Australian 
Telescope's f/3.3 Tek 1K$^{2}$ 24$\mu m$ pixel CCD (pixel size 0.39 by 0.39 
arcsecs, f.o.v. 6.5 by 6.5 arcmins), but should be applicable to any similar
field-of-view detector. The primary results for this setup are summarised as 
follows:

~

(i)\,the cluster richness must be greater than 1.5,

(ii)\,the cluster should lie in the redshift range $0.1 < z < 0.3$,

(iii)\,the FWHM of the seeing must be less than 1.25$''$.

(iv)\,the technique is independent of cluster type ({\it i.e.} morphological 
mix or profile shape) for the shapes and mixes explored.

(v)\,the technique can recover the faint end slope of the cluster LD for any 
shape distribution

(vi)\,photometric zero points must be accurate to $\Delta m = \pm 0.12$.

~

The simulations have demonstrated that for a given detector and exposure time
there is a range of observable parameter space over which the cluster LD 
reconstruction method is valid and beyond which results are questionable. 
Detailed and exhaustive simulations such as those presented here are therefore 
vital and essential to establish the credibility of this technique for any 
specified dataset. Within the valid boundaries the technique is extremely time 
efficient, simple and likely to make a substantial contribution to the study 
of cluster populations over a range of redshift and environment.
Finally we note that the simulation software can generate data for any 
detection setup, field-size etc for cluster or field images.
Other groups wishing to use this software are encouraged and should contact
spd@edwin.phys.unsw.edu.au

\section*{Acknowledgments}
We would like to thank Stuart Ryder and Paul Bristow for useful discussions. 
SPD and WJC acknowledge the financial support of the Australian Research 
Council. SP is
supported by a Royal Society University Fellowship. We also thank Emmanuel
Bertin for making his SExtractor software publicly 
available, and SUN Microsystems for their continued support of the
Astrophysics group at UNSW.

\section*{References}

\small

\begin{description}

\item Abell G.O., 1958, ApJ. Supp., 3, 211

\item Abell G.O., Corwin Jr H.G., Olowin R.P., 1989, ApJ. Supp., 70, 1

\item Bernstein G.M., Nichol R.C., Tyson J.A., Ulmer M.P., Wittman D., 1995, 
AJ, 110, 1507

\item Bertin E., Arnout, S., 1996, Astr. Astrophys. Suppl., 117, 393

\item Binggeli B., 1994, in ESO/OHP Workshop proceedings on
{\it Dwarf Galaxies}, eds G. Meylan and P. Prugniel (ESO)

\item Biviano A., Durret F., Gerbal D., Le Fevre O., Lobo C., Mazure A.,
Slezak E., 1995, A \& A, 297, 610

\item Boroson T., 1981, Ap J Supp, 46, 177

\item Broadhurst T., 1998, ApJL, submitted (astro/ph9511150)

\item Coleman, G.D., Wu, C.C., Weedman, D.N., 1980, Ap J Supp, 43, 393

\item de Propris R., Pritchet C.J., Harris W.E., McClure R.D., 1995, ApJ, 450,
534

\item de Vaucouleurs G., 1959, Handbuch der Physik, 53, 311

\item Disney M.J., 1995, in ``Opacity of Spiral Galaxies'', eds Davies J.I., 
and Burstein, D., (Dordrecht, Kluwer), p 5

\item Dressler, A., 1980, Ap J, 236, 351

\item Driver, S.P., Couch W.J., Phillipps S., 1998, MNRAS, in press (Paper III)

\item Driver, S. P., Phillipps, S., Davies, J. I., Morgan, I., 
Disney, M.J., 1994a, MNRAS, 268, 393

\item Driver, S. P., Phillipps, S., Davies, J. I., Morgan, I., 
Disney, M.J., 1994b, MNRAS, 266, 155

\item Driver, S. P., Windhorst, R. A., Griffiths R. E. 1995, ApJ,
453, 48

\item Driver, S.P., Windhorst, R. A., Ostrander, E.J., Keel W.C.,
Griffiths, R. E., Ratnatunga, K.U., 1995, ApJL, 449, L23

\item Driver, S.P., Phillipps, S., 1996, ApJ, 469, 529

\item Drinkwater, M.J., Currie M.J., Young C.K., Hardy E., Yearsley J.M.,
1996, MNRAS, 279, 595

\item Ferguson, H.C., Binggeli., 1994, A\&ARv, 6, 67

\item Freeman K.C., 1970, ApJ, 160, 811

\item Glazebrook, K., Ellis, R. S., Colless, M. M., Broadhurst, T. J.,
Allington-Smith, J., Tanvir, N., 1995, MNRAS, 273, 157

\item Glazebrook, K., Ellis, R. S., Santiago, B., Griffiths, R.E., 
1995, MNRAS, 275, 19pp

\item Holmberg E., 1946, Medd. Lunds Obs II, No 117

\item Hubble E., 1926, ApJ, 64, 321

\item Impey C.D., Bothun G.D., Malin D.F., 1988, Ap J, 330, 634

\item King C.J., Ellis, R.S., 1985, ApJ, 288, 456

\item Kron R.G., 1978, PhD Thesis, University of California at Berkeley

\item Marzke, R.O., Geller, M.J., Huchra, J.P., Corwin Jr, H.G., 1994, 
A J, 108, 437

\item Metcalfe, N., Shanks, T., Fong, R., Roche, N., 1995, MNRAS, 273, 257

\item Odewahn S.C., Windhorst R.A., Driver S.P., Keel W.C., 1996, ApJ, 472, L13

\item Oemler Jr, A., 1974, Ap J, 194, 1

\item Pence W., 1976, Ap J, 203, 39

\item Phillipps, S., Davies, J.I., Disney, M.J., 1990, MNRAS, 242, 235

\item Phillipps, S., Driver, S.P., 1995, MNRAS, 274, 832 

\item Sandage, A., 1961, Ap J, 133, 355

\item Schechter, P., 1976, Ap J, 203, 297

\item Simien, F., de Vaucouleurs, G., 1986, Ap J, 302, 564

\item Smith, R.M., Driver, S.P., Phillipps, S., 1997, MNRAS, 287, 415 (Paper I)

\item Trentham N., 1997a, MNRAS, 286, 133

\item Trentham N., 1997b, MNRAS, 290, 334

\item Trentham N., 1998, MNRAS, 295, 360

\item Wall J.V., Quarterly Journal of the R.A.S., 1996, 37, 519

\item Williams R.E., {\it et al.} 1996, AJ, 112, 1335 

\item Wilson, G, Smail, I., Ellis, R.S., Couch, W.J., 1997, MNRAS, 284, 915 

\end{description}

\normalsize

\appendix

\section{Simulation details}
The description in \S 2 defines the generic galaxy types which we shall 
consider. Specific details ({\it i.e.} actual numbers) have been listed in 
Table 1. The most important {\it a priori} decision was the opacity. While 
detailed models are available to cope with opacity, here we decide
to assume all objects are optically thin at all radii 
({\it i.e.} zero opacity). The main motivation for this, apart from the fact 
that the true opacity of disk 
systems is not known (c.f. Disney 1995), is that the majority
of objects simulated will be barely resolved and the additional computational 
time required to compute the effect of opacity was not justified. 
Setting opacity to zero also simplifies the simulation process as the apparent 
magnitude of an object remains fixed regardless of inclination. This in turn, 
allows us to fully define a galaxies light-profile simply from its absolute 
magnitude ($M$), intrinsic surface brightness ($\mu_{o}$), and apparent 
magnitude ($m$) (plus a randomly selected orientation/ellipticity).

\subsection{Cosmological Model}
Our software, which generates the simulated profiles, uses the output from
our number-count software (c.f. Driver, Windhorst \& Griffiths 1995) which 
(taking a normalisation parameter 
for each LF type) generates a list of number-counts, N(M,m,T). The 
specification of type (T) and absolute magnitude also 
defines the intrinsic face-on surface brightness ($\mu_{o}$, 
or in intensity units $I_{o}$), if we 
assume the relationships of Table 1 (taken from Binggeli 1994). By assuming
zero opacity, the redshift ($z$) can be derived via the relevant 
distance-luminosity relation only ({\it i.e.} without requiring knowledge of 
inclination etc). Here the cosmological model we have chosen
is the standard model of a flat expanding universe, with 
$H_{0}$ = 50 km s$^{-1}$ Mpc$^{-1}$, $\Omega$ = 1, $\Lambda$ = 0, 
$q_{0}$ = 0.5. The effects of such a model on the visibility of galaxies is 
discussed in Phillipps, Davies \& Disney (1990) and the distance luminosity 
relation is defined as:
\begin{equation}
M=m-5\log{d_{l}}-25-k_{c}(z),
\end{equation}
\noindent where $d_{l}$, the luminosity distance, is dependent on the World 
model:
\begin{equation}
d_{l}=(1+z)2cHo^{-1}[1-(1+z)^{-1/2}],
\end{equation}
and $k_{c}(z)$ is the K-correction.

\vspace{0.5cm}

Evaluating the above gives $z$ and within this cosmological model the apparent 
($I_{o}^{app}$) and intrinsic ($I_{o}$) face-on 
central surface intensities are related by:
\begin{equation}
I^{app}_{o} = \frac{I_{o}10^{-0.4k_{c}(z)}}{(1+z)^{4}}.
\end{equation}
This gives us the apparent face-on central surface brightness of the object
which, with the known apparent magnitude (m) and type, fully defines the light 
profile.

\subsection{Light Profiles}
The basic light-profile for all galaxy types is assumed to follow the 
generalized form:
\begin{equation}
I(r) = I_{o} exp \left(- \left(\frac{r}{a_{o}} 
\right)^{\frac{1}{\beta}}\right)
\end{equation}
\noindent where,
$I(r)$ is the intensity at a radius $r$,
$a_{o}$ is the face-on scale-length,
$\beta = 1$ for pure disks, (Freeman 1970) and $\beta = 4$ for ellipticals 
(de Vaucouleurs 1959).

\vspace{0.5cm}

\noindent Integrating this light profile over a galaxy's total {\it face-on} 
area gives the total luminosity:
\begin{equation}
L_{Total} = (2 \beta) ! \pi I_{o} a_{o}^{2}
\end{equation}
or in magnitude units;
\begin{equation}
m = \mu^{app}_{o} + 5 \log(a_{o}) + 2.5 log ((2\beta)!\pi) 
\end{equation}
Hence as m, $\mu^{app}_{o}$ and $\beta$ are already known, $a_{o}$ is
defined.

\vspace{0.5cm}

The final stage is the randomisation of orientation. For ellipticals we simply
allocate an ellipticity assuming equal probability for all ellipticities from
E0 to E7 and define the product of the major and minor axis to equal $a_{o}$,
the E0 scale-length. For other types we allocate an inclination angle 
$\theta$ in the range 0---90 degrees and compute the ellipticity as described 
by Eqn 1, (c.f \S 2.1). Hence
the surface brightness of ellipticals is constant regardless of
ellipticity, however the central surface brightness of disks is
dependent on inclination as: 
$\mu^{app}_{o} = \mu^{app}_{i} - 2.5 \log (cos \theta)$
Boroson (1981)\footnote{Strictly speaking this assumes a flat luminosity 
profile, again a somewhat over simplification, 
for a more detailed discussion of surface-brightness inclination 
effects (see Disney, Davies \& Phillipps 1990).}. Finally the 
elliptical light-profile is simulated by randomly selecting a position
angle and converting the circular intensity light-profile (Eqn 6) to that for 
an ellipse:
\begin{equation}
I(r,PA) = I^{app}_{i} 
\exp 
\left( - 
\left( 
\frac{
r \sqrt{a^{2} cos^{2}(PA) + b^{2} sin^{2}(PA)}
}
{a^{app}_{i}}
\right)^{\frac{1}{\beta}} \right)
\end{equation}
where a and b are the major and minor axis with their product equal to unity.

\pagebreak

\section*{Tables}

\begin{table}[h]
\footnotesize
\caption{The intrinsic parameters and assumptions used for the simulations 
for each of the four generic galaxy types.}

~

\begin{tabular} {lllll} \hline \hline
Property      & elliptical & mid-type spiral & irregular & dEs \\ \hline
$M^{*}_{R}$   & $-22.3$    & $-21.7$         &  $-21.2$  & N/A  \\
$\alpha$      & $-0.9$     & $-0.8$          &  $-1.8$   & N/A  \\
$\phi_{*}$    & $1.14 \times 10^{-3}$ & $1.74 \times 10^{-3}$ &
$2.50 \times 10^{-4}$ & N/A  \\
Profile       & $I_{r}=I_{o}exp(-\frac{r}{a}^{\frac{1}{4}})$ &
$I_{r^{Bulge}}=I_{o}exp(-\frac{r}{a}^{\frac{1}{4}})$ &
$I_{r}^{Global}=I_{o}exp(-\frac{r}{a})$ & $I_{r} =I_{o}exp(-\frac{r}{a})$ \\
              &            &
$I_{r}^{Disk}=I_{o}exp(-\frac{r}{a})$ &
$I_{r}^{Local}=0.05I_{o}exp(-\frac{r}{a})$ & \\
$\mu_{o}$    & 
$\mu_{R}$=$17.6-(M_{R}+26.4)$ &
$\mu_{R}$=19.3 &
$\mu_{R}$=19.8+0.8($M_{R}$+18.9) &
$\mu_{R}$=16.6+($M_{R}$+20.4) \\
$(B-R)$ & 1.8 & 1.4 & 0.9 & 1.8 \\
Ellipticity & 0.0---0.7 & 0.0---0.8 & 0.0---0.5 & 0.0---0.3 \\
K$_{R}$-corr & 
$1.36z+1.07z^{2}$ & 
$0.40z+0.71z^{2}$ &
$0.05z+0.78z^{2}$ &
$1.36z+1.07z^{2}$ \\ \hline
\end{tabular} 

\vspace{0.25cm}

\noindent
The parameters for this table are taken from the following sources:
$M_{R}^{*}$ \& $\alpha$, Marzke {\it et al.} (1995);
$\phi_{*}$, Driver {\it et al.} (1995);
$Profiles$, Freeman (1970) for disks and de Vaucouleurs (1959) for ellipticals;
$\mu_{o}$, Binggeli (1994);
$(B-R)$, Coleman, Wu \& Weedman (1980);
$Ellipticity$, Sandage (1961) and Hubble (1926);
$K-corrections$, Driver {\it et al.} (1994b, and originally from Pence 1976 
and King \& Ellis 1985).

\normalsize

\end{table}

\begin{table}[h]
\caption{The measured flux for the four types of simulated galaxy as a 
function of aperture size (defined in terms of Kron Radii). 
All galaxies were simulated to have $m_{R} = 18.0$ and
the tabulated values are the median measured value for 54 randomly simulated
examples for each type.}

\vspace{0.5cm}

\begin{tabular}{ccccc} \hline \hline
Aperture Radius & E & Sb & Irr & dE \\ \hline
2.0$R_{K}$ & 18.16 & 18.05 & 18.01 & 18.02 \\
2.5$R_{K}$ & 18.16 & 18.05 & 18.01 & 18.02 \\
3.0$R_{K}$ & 18.14 & 18.04 & 18.00 & 18.01 \\
3.5$R_{K}$ & 18.11 & 18.02 & 18.00 & 18.01 \\
4.0$R_{K}$ & 18.09 & 18.02 & 18.00 & 18.00 \\
4.5$R_{K}$ & 18.07 & 18.01 & 18.00 & 18.00 \\
5.0$R_{K}$ & 18.06 & 18.01 & 18.00 & 18.00 \\ \hline
\end{tabular}
\end{table}

\pagebreak

\begin{figure}[p]

\vspace{-3.0cm}

\centerline{\hspace{-7.0cm} \psfig{file=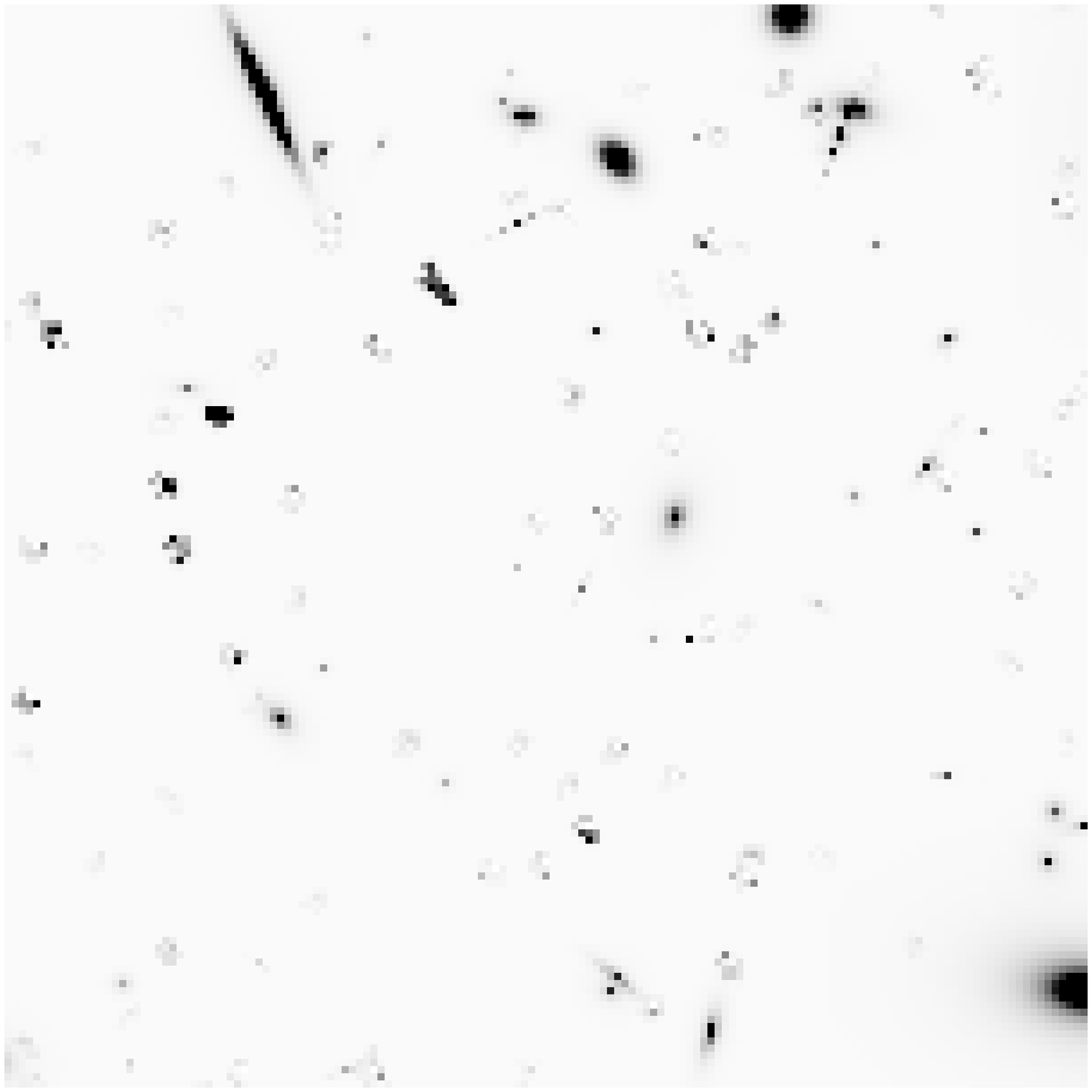,height=80mm,width=60mm}}

\vspace{-8.0cm}

\centerline{\hspace{7.0cm} \psfig{file=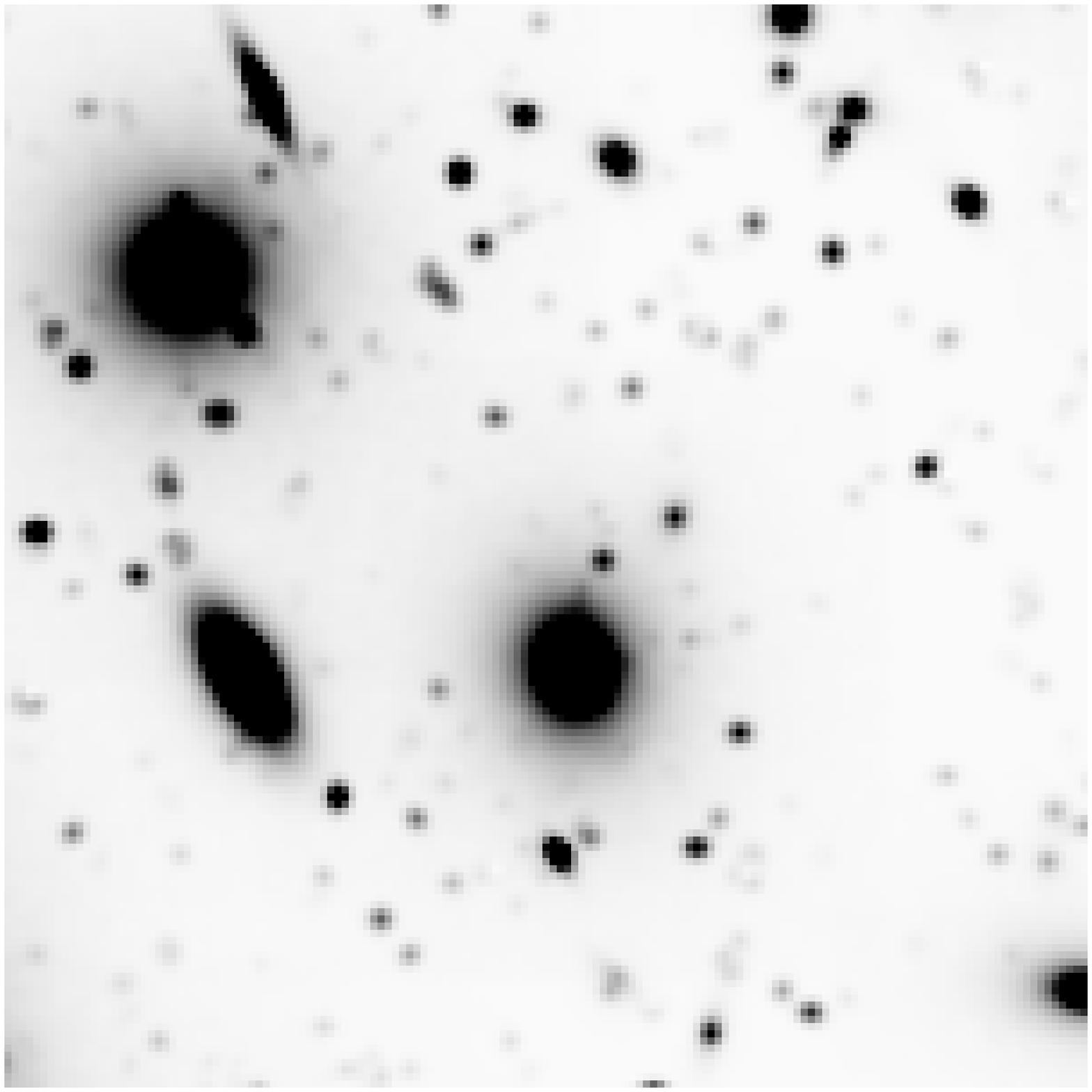,height=80mm,width=60mm}}

\vspace{-0.75cm}

\hspace{2.0cm} (a) Field galaxies \hspace{4.0cm} (d) Convolved with Gaussian

\vspace{-0.25cm}

\centerline{\hspace{-7.0cm} \psfig{file=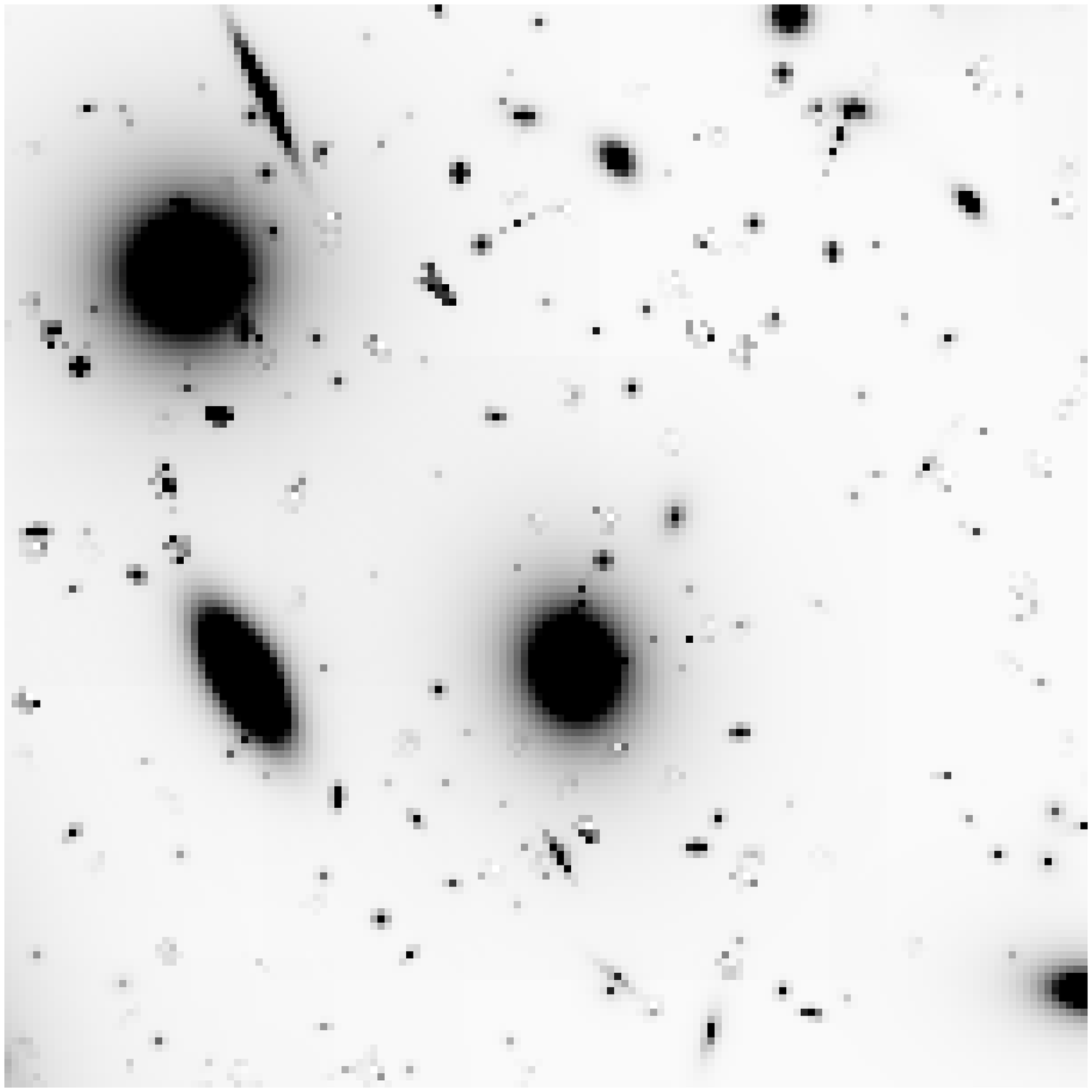,height=80mm,width=60mm}}

\vspace{-8.0cm}

\centerline{\hspace{7.0cm} \psfig{file=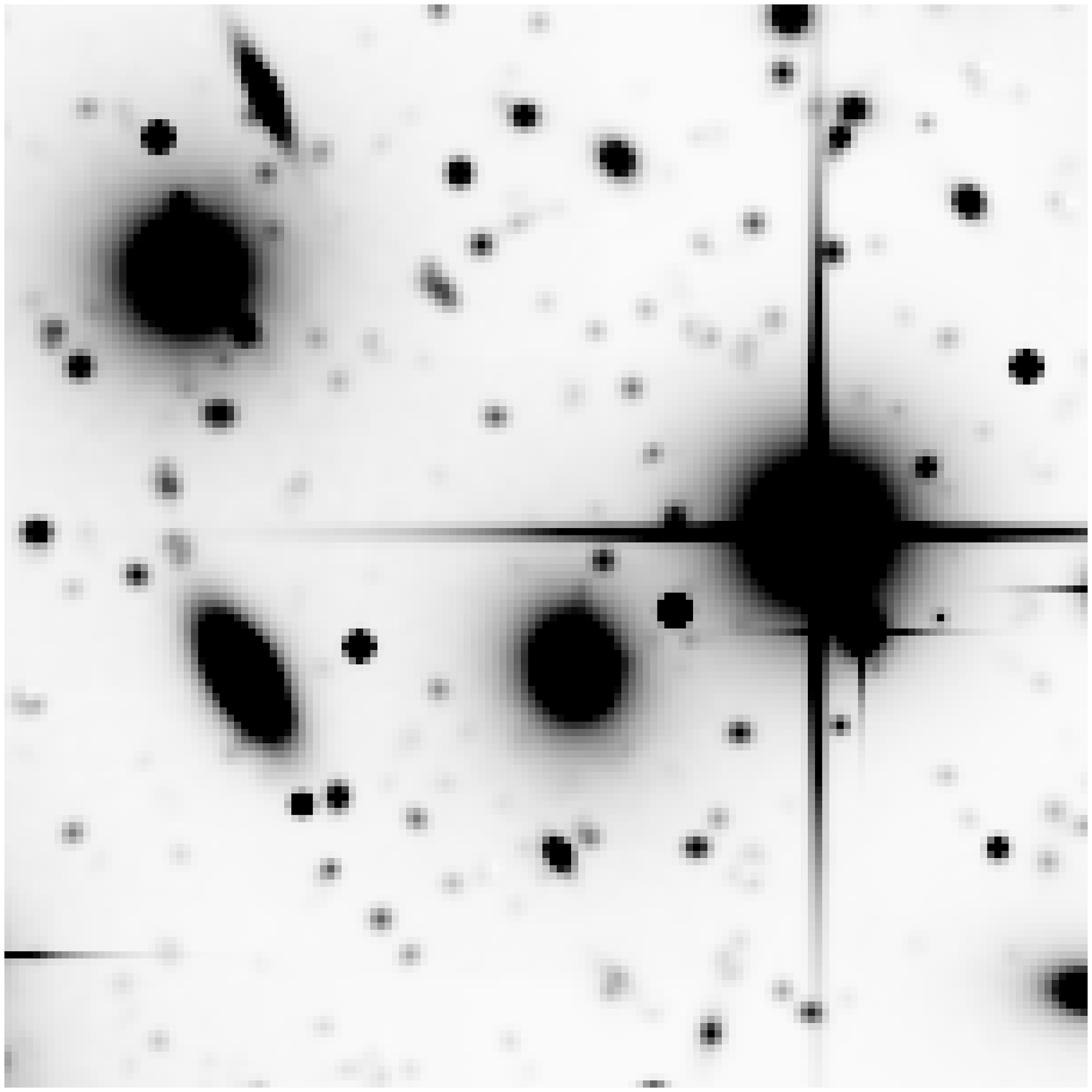,height=80mm,width=60mm}}

\vspace{-0.75cm}

\hspace{2.0cm} (b) Cluster added \hspace{4.0cm} (e) Stars added

\vspace{-0.25cm}

\centerline{\hspace{-7.0cm} \psfig{file=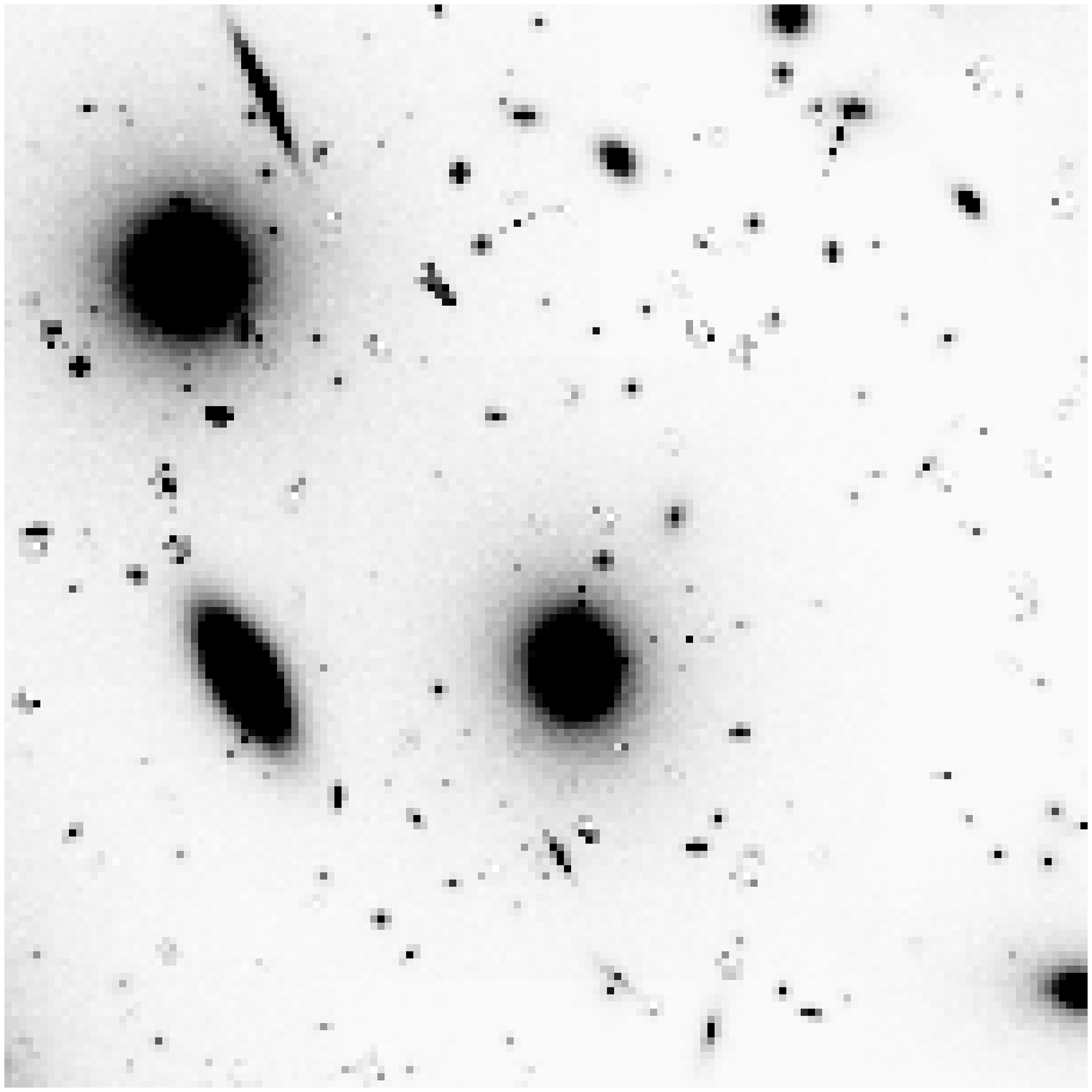,height=80mm,width=60mm}}

\vspace{-8.0cm}

\centerline{\hspace{7.0cm} \psfig{file=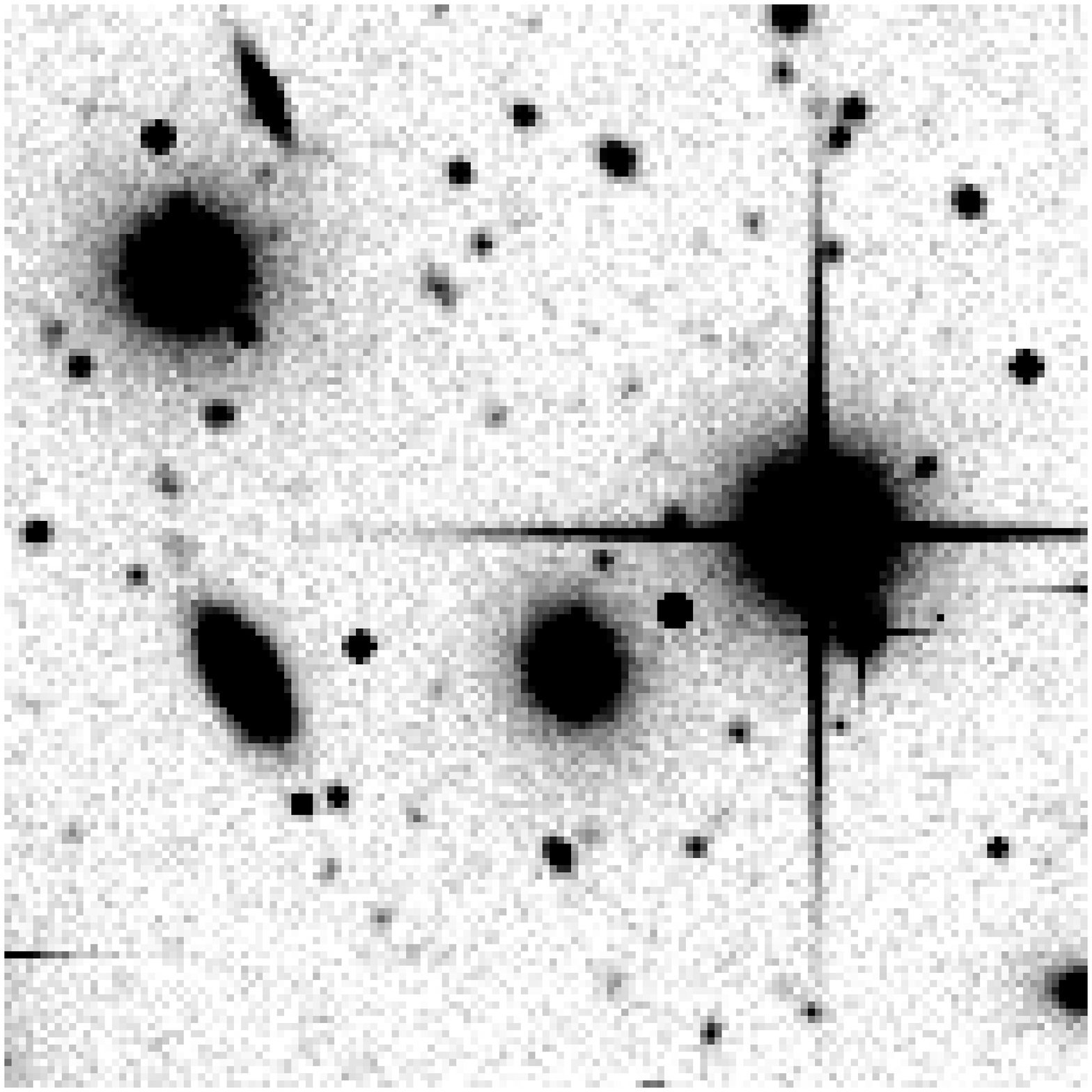,height=80mm,width=60mm}}

\vspace{-0.75cm}

\hspace{2.0cm} (c) Shot Noise added \hspace{3.5cm} (f) Sky noise added

\caption{An illustration of the process of simulating a section of a sight-line
through a cluster. The simulated field of view is 1 sq arcmin, pixels size is 
0.39 by 0.39 arcsecs and the seeing FWHM is 0.9$''$. }

\end{figure}

\begin{figure}[p]
\vspace{-1.5cm}
\centerline{\hspace{0.0cm} \psfig{file=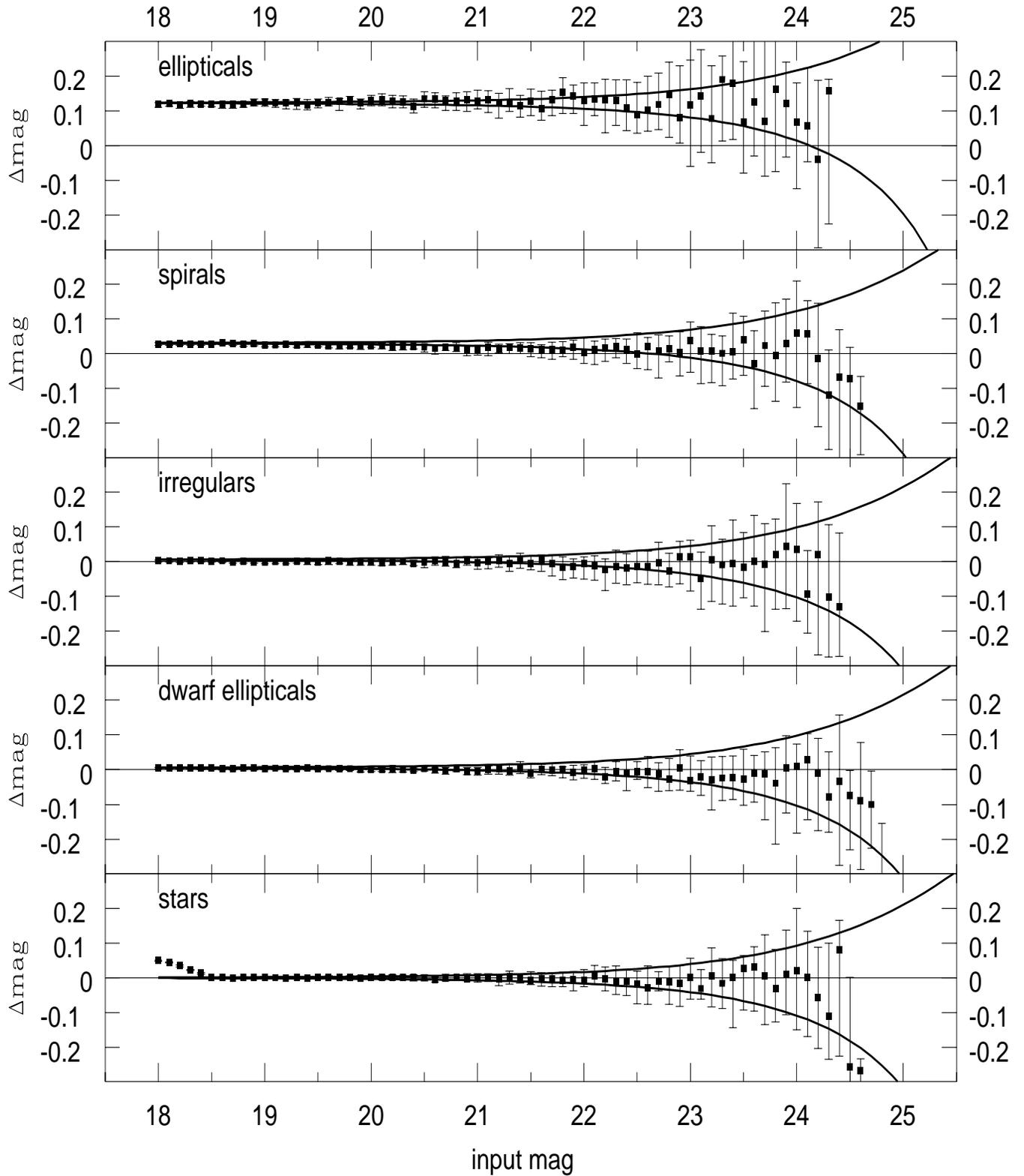,height=220mm,width=180mm}}
\vspace{-0.5cm}
\caption{The difference between the measured magnitude and allocated magnitude 
for the four types of simulated galaxies and simulated stars. The solid line 
indicates the zero discrepancy line. The thick lines indicate the theoretical 
upper and lower quartiles based on the adopted sky noise.
54 isolated objects at random orientations were 
simulated from $m_{R} = 18.0$ to $m_{R} = 26$ at 0.1 magnitude intervals.}
\end{figure}

\begin{figure}[p]
\vspace{-1.5cm}
\centerline{\hspace{0.0cm} \psfig{file=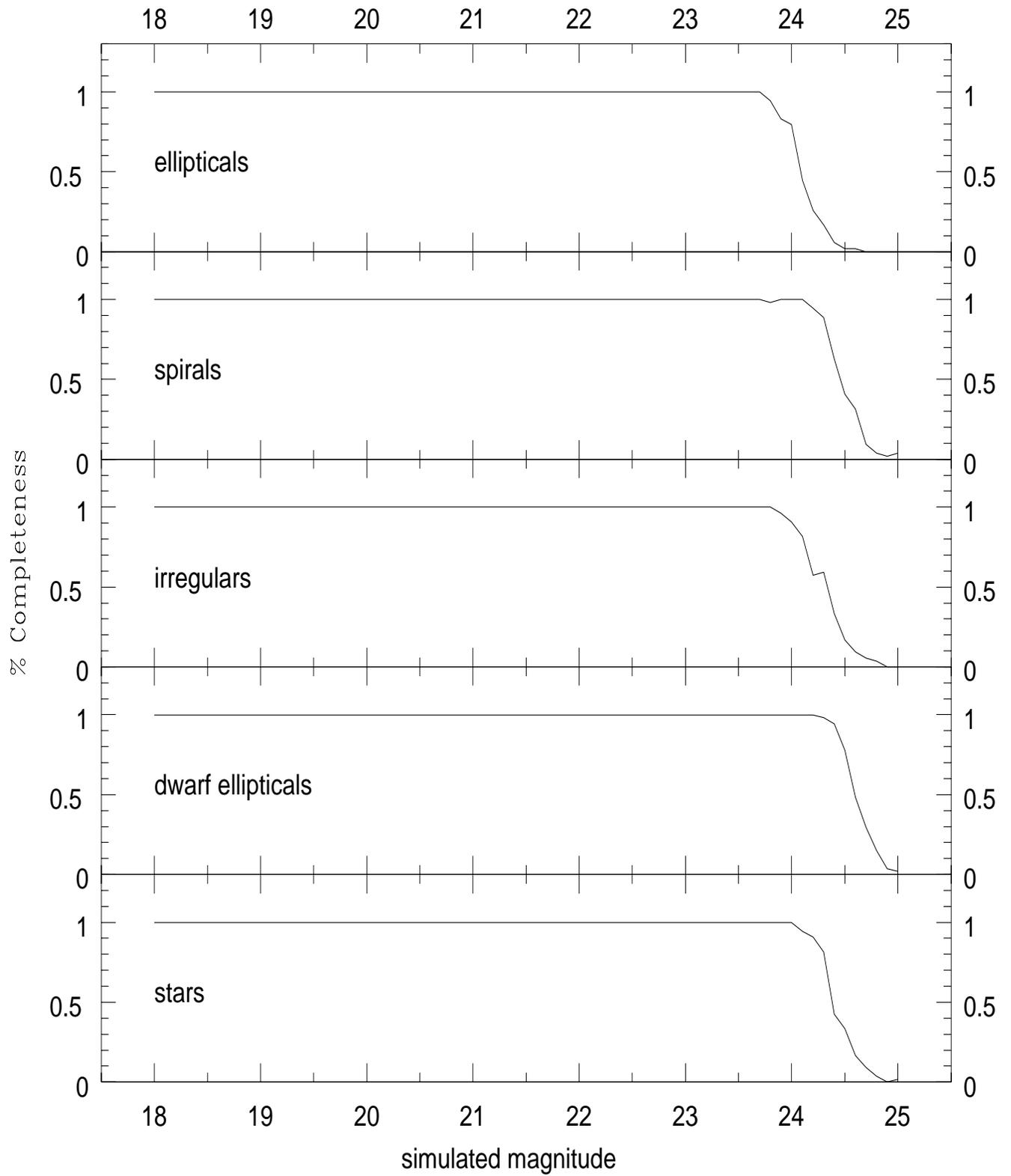,height=220mm,width=180mm}}
\caption{The fractional completeness for the simulations shown in Figure 2, as 
a function of input magnitude.}
\end{figure}

\begin{figure}[p]
\vspace{-1.5cm}
\centerline{\hspace{0.0cm} \psfig{file=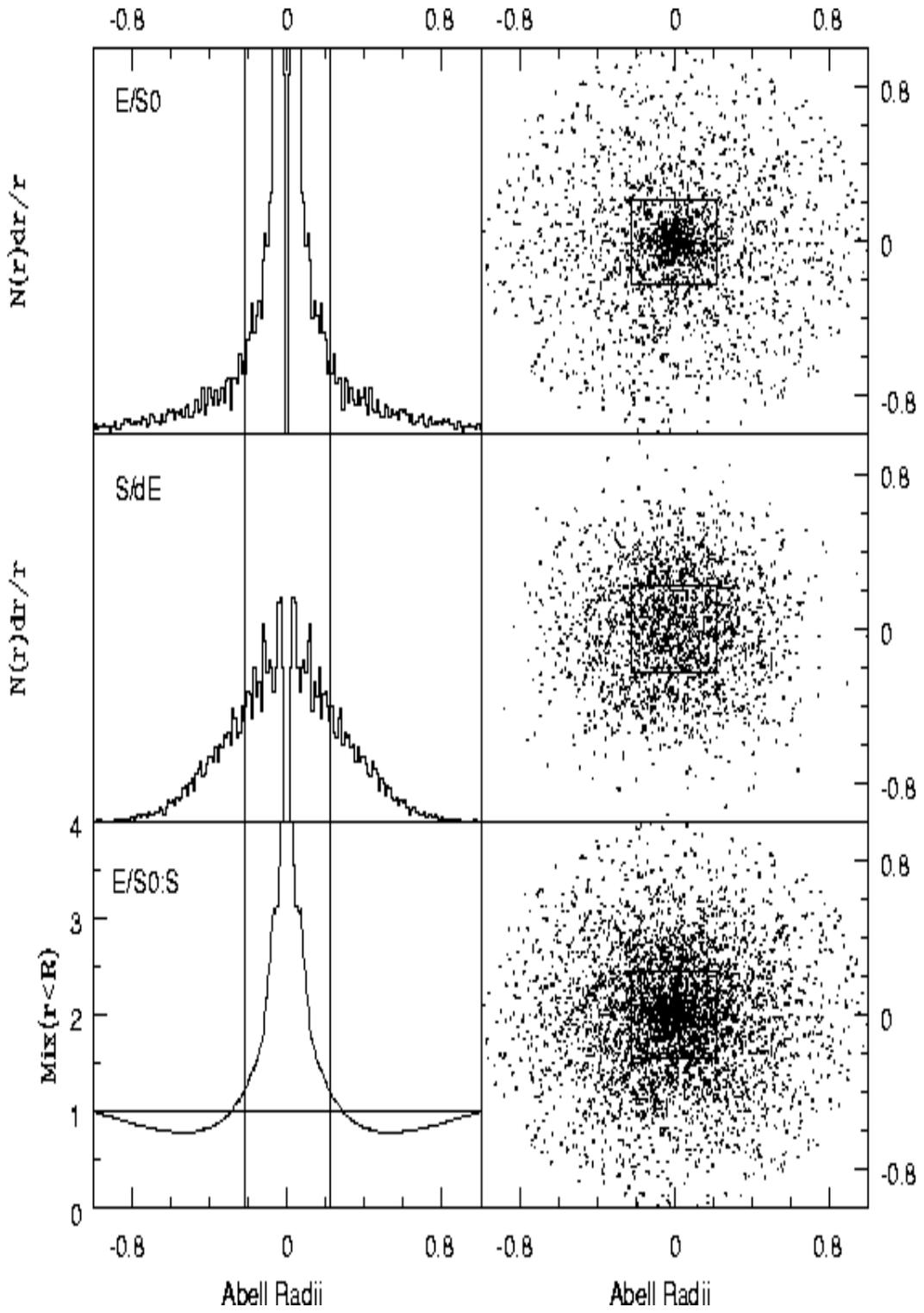,height=200mm,width=140mm}}
\caption{The density profiles for ellipticals (upper left), for spirals and
for dwarf ellipticals (middle left) and the morphological mix of E/S0:Spirals
as a function of Abell radius (lower left), the square box within the panels 
represents the equivalent field of view for the observational data presented in
Paper III. The right panel shows the
projected density of elipticals (upper right), spirals (middle right)
and ellipticals and spirals (lower right).}
\end{figure}

\begin{figure}[p]
\vspace{-3.0cm}
\centerline{\hspace{-1.0cm} \psfig{file=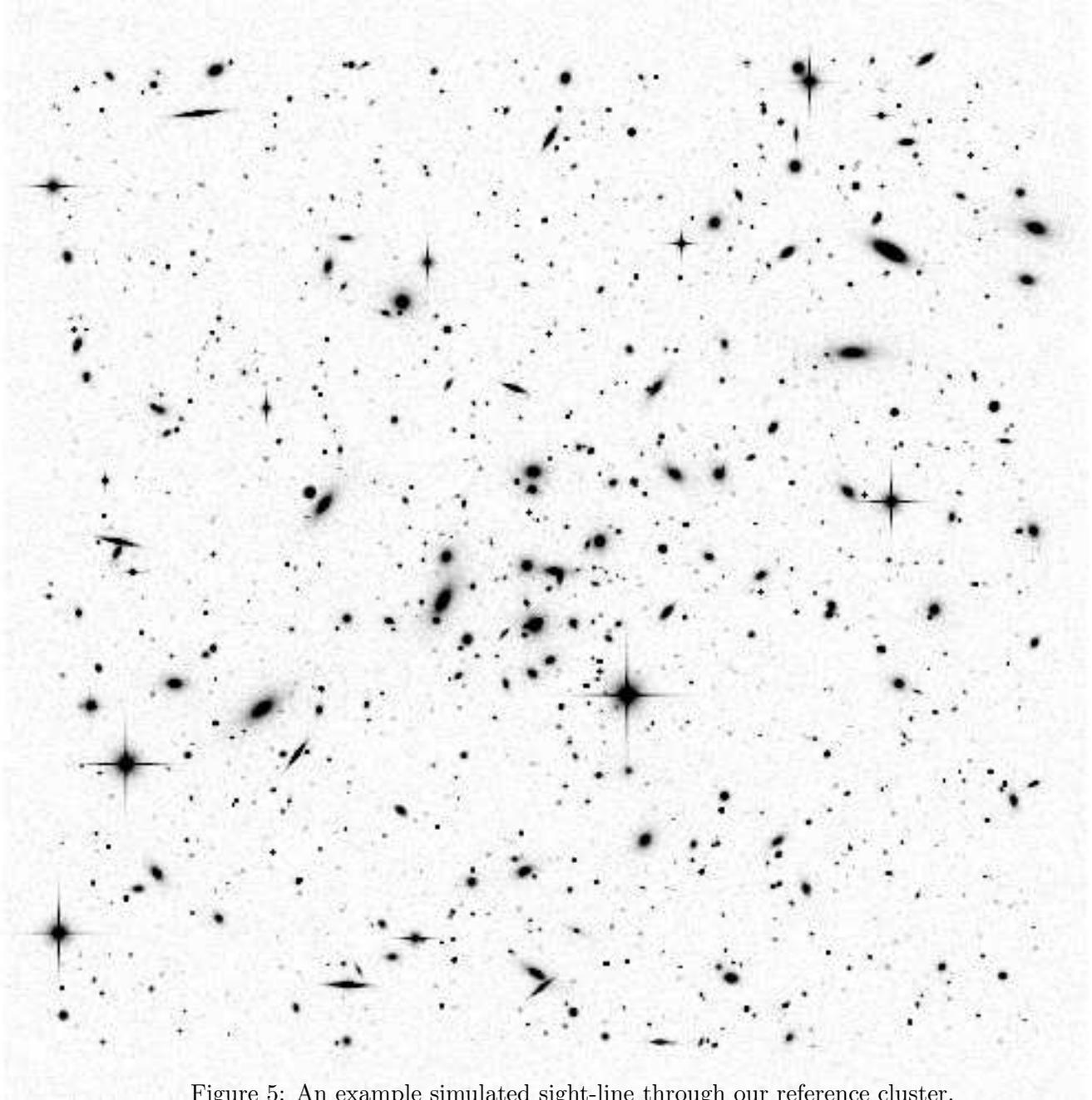}}
\vspace{-1.0cm}
\caption{An example simulated sight-line through our reference cluster.}
\end{figure}

\begin{figure}[p]
\vspace{-1.5cm}
\centerline{\hspace{0.0cm} \psfig{file=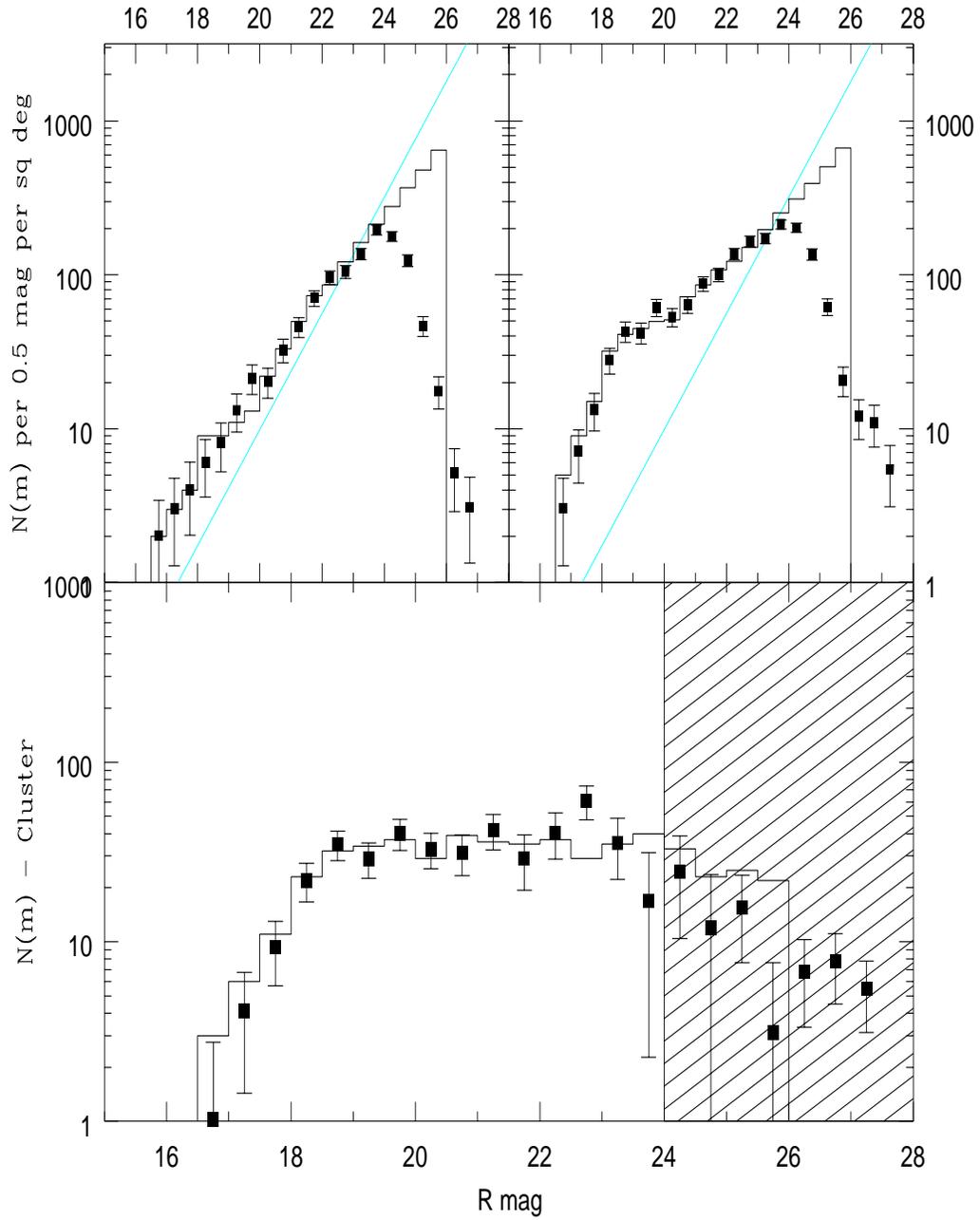,height=180mm,width=135mm}}
\caption{The number-count plots for the comparison field (upper left),
cluster sight-line (upper right) and the excess as compared to 
the known input distributions (lower panel). The solid 
straight lines show the mean
galaxy counts from Metcalfe {\it et al.} (1996) for reference.}
\end{figure}

\begin{figure}[p]
\vspace{-1.5cm}
\centerline{\hspace{0.0cm} \psfig{file=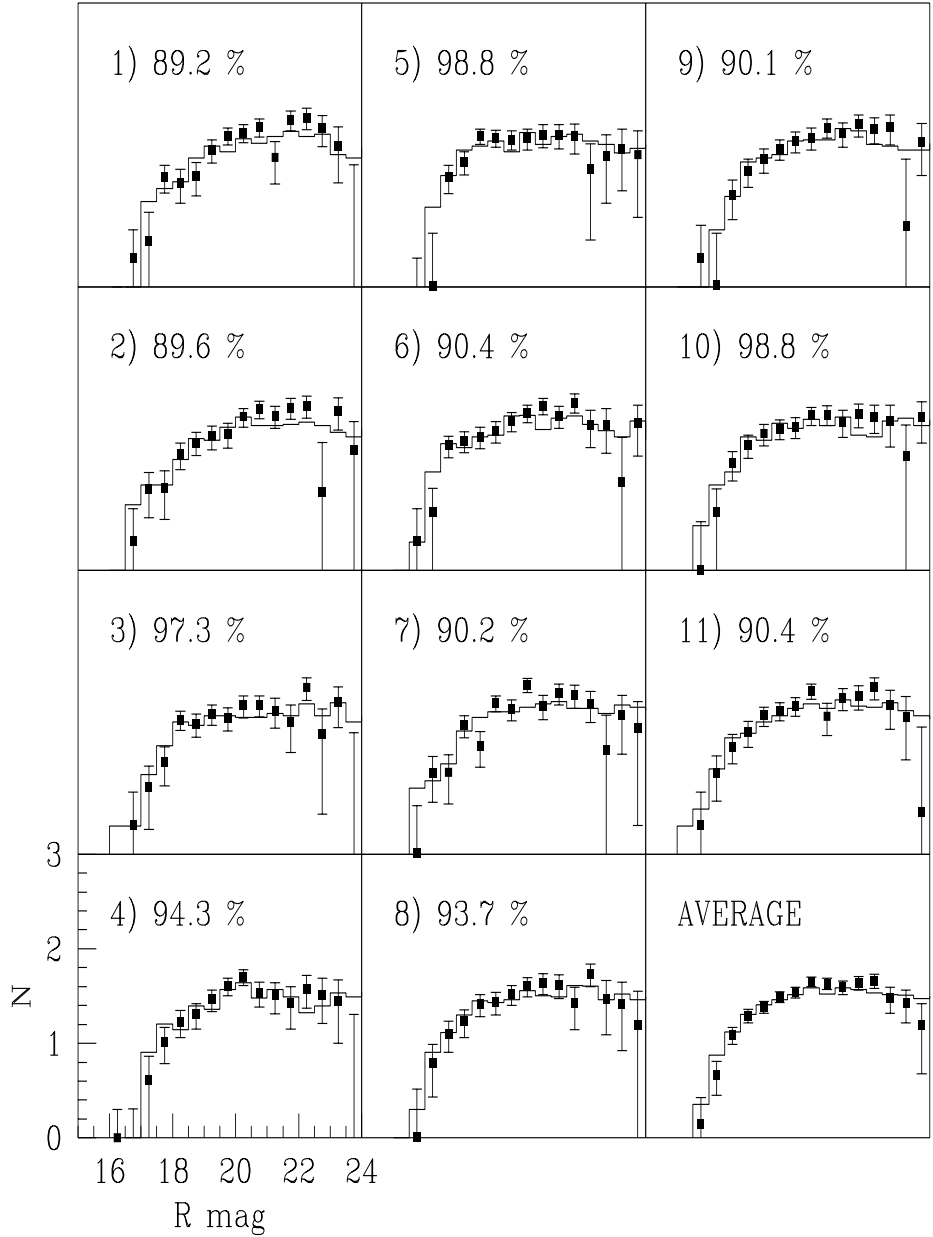,height=180mm,width=135mm}}
\caption{The recovered LDs compared to input LDs for 11 realisations of our
reference cluster. The bottom right panel shows the average results.}
\end{figure}

\begin{figure}[p]
\vspace{-1.5cm}
\centerline{\hspace{0.0cm} \psfig{file=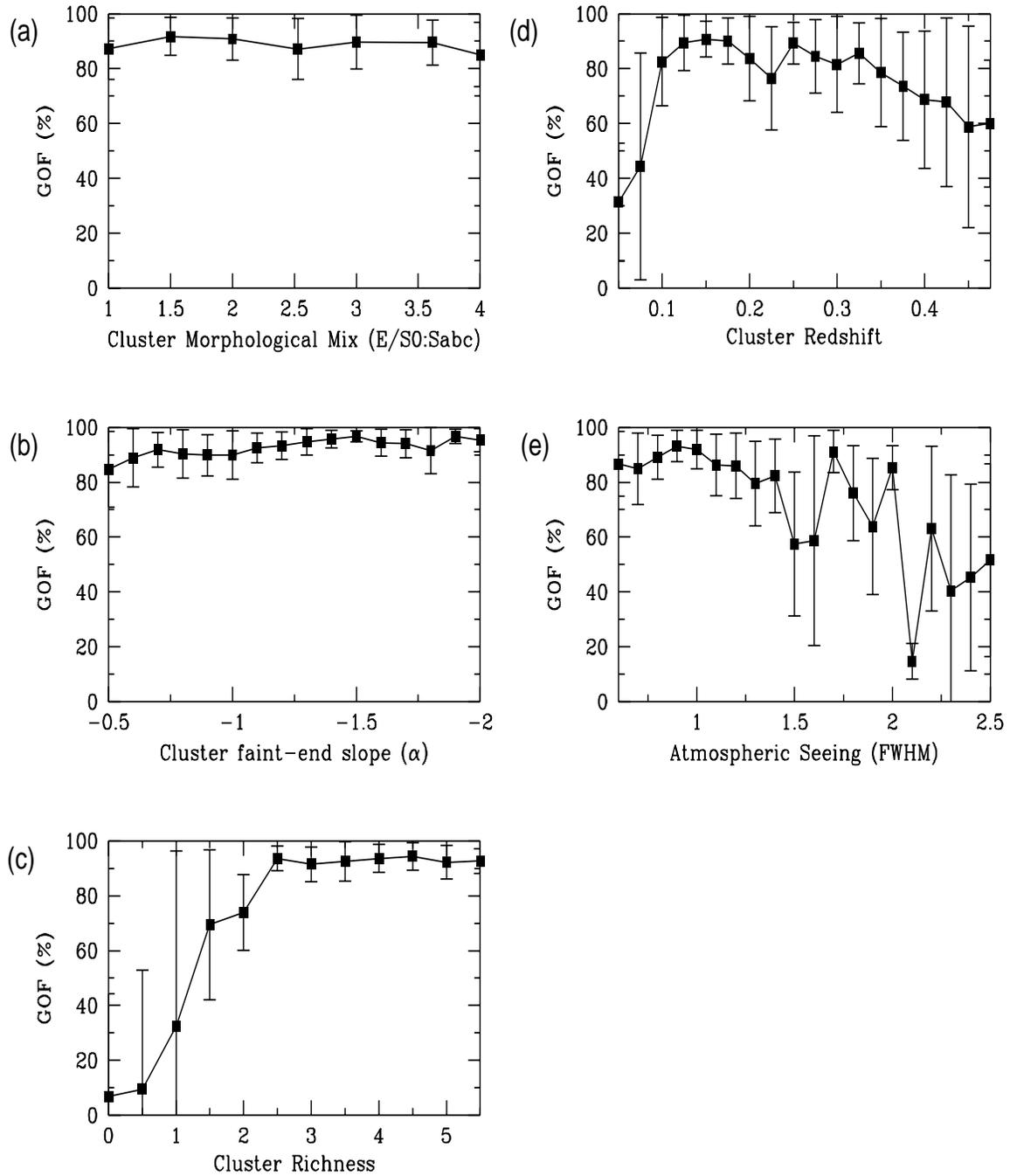,height=200mm,width=160mm}}
\caption{Simulation results investigating the method's dependence on: (a) 
morphological mix, (b) faint end slope, (c) richness, (d) redshift and (e) 
seeing.}
\end{figure}

\end{document}